\documentclass{article}
\usepackage{fullpage,amsmath,amssymb,amsthm,citesort,spconf}
\usepackage{epsfig,graphicx,subfigure,wrapfig,psfrag}		
\graphicspath{{./figs}}

\def \reals {\mathbb{R}}

\newcommand{\E}[2]{\mathbb{E}_{#2}\left\{ #1 \right\}}
\newcommand{\Proba}[2]{\mathbb{P}_{#2}\left\{ #1 \right\}}

\newtheorem{thm}{Theorem}[section]

\title{Multi-target Radar Detection within a Sparsity Framework}
\name{Han Lun Yap$^{1}$ and Radmila Pribi\'{c}$^{2}$}

\address {$^1$Sensors Division, DSO National Laboratories, Singapore\\$^2$Sensors Advanced Developments, Thales Nederland Delft, The Netherlands}

\begin{document}

\maketitle

\begin{abstract}
Traditional radar detection schemes are typically studied for single target scenarios and they can be non-optimal when there are multiple targets in the scene. 
In this paper, we develop a framework to discuss multi-target detection schemes with sparse reconstruction techniques that is based on the Neyman-Pearson criterion. 
We will describe an initial result in this framework concerning false alarm probability with LASSO as the sparse reconstruction technique. 
Then, several simulations validating this result will be discussed. 
Finally, we describe several research avenues to further pursue this framework. 
\end{abstract}

\section{Introduction}

\begin{figure}[t]
	\begin{center}
\includegraphics[width=8cm]{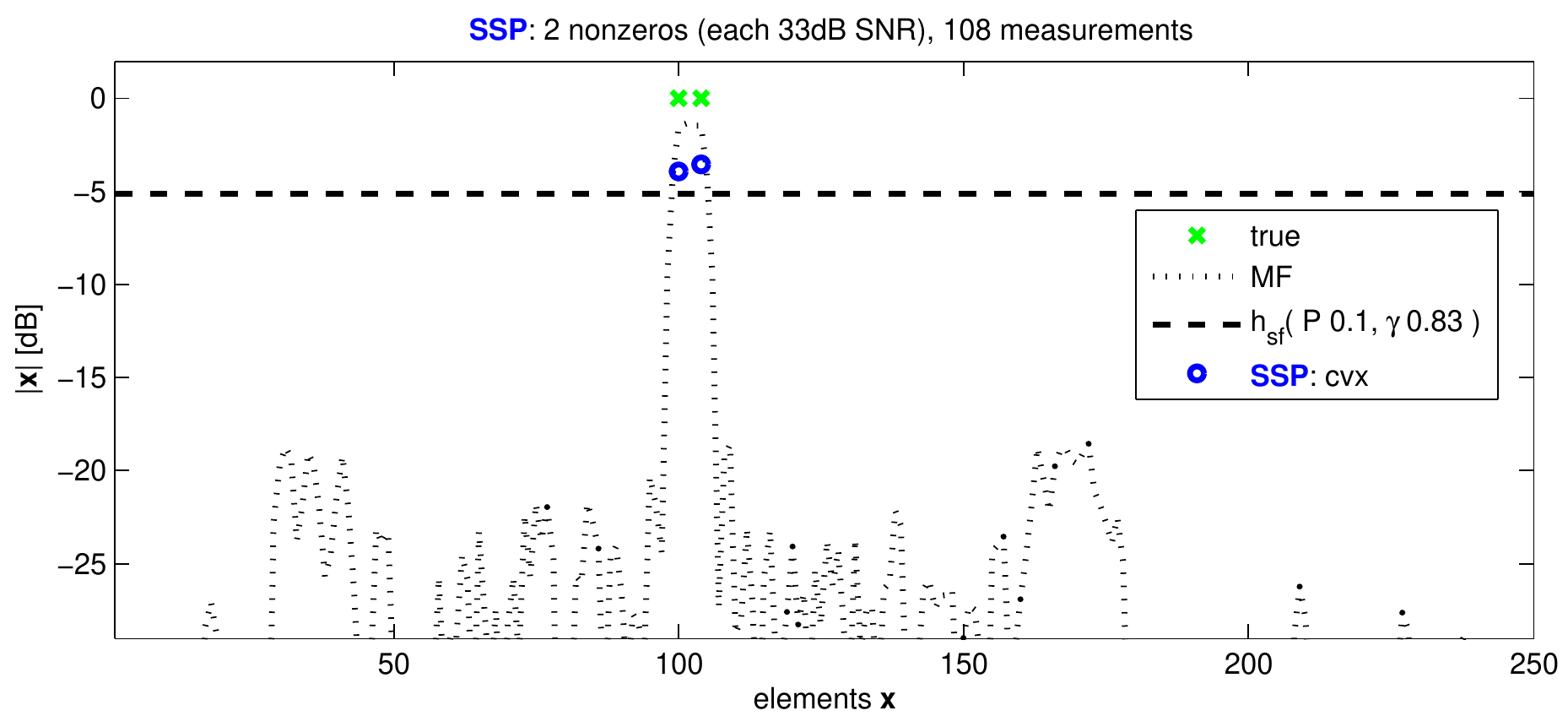} \\
		\includegraphics[width=8cm]{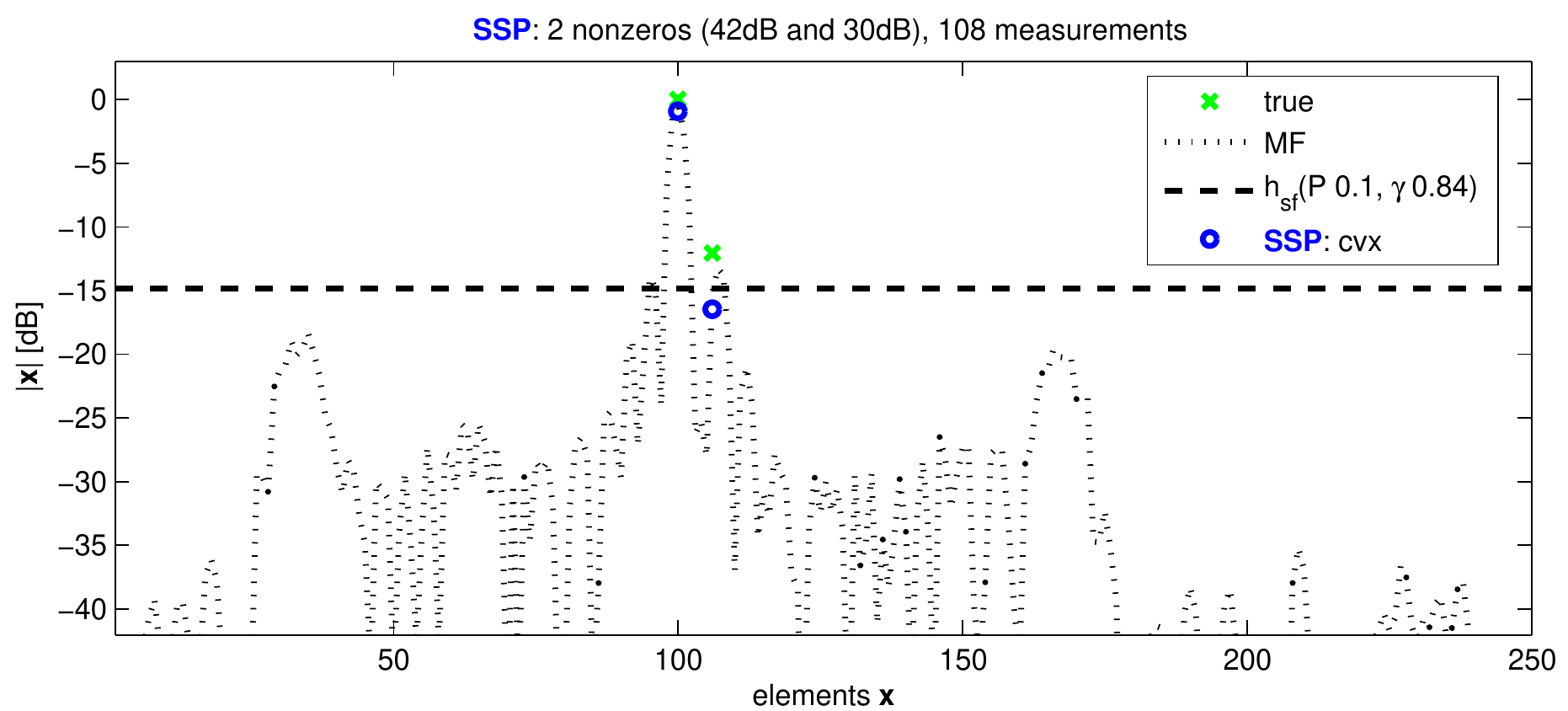}
	\end{center}
	\vspace{-5mm}
	\caption{\small(a) Two targets (crosses) close together in range space $x$ that are unresolved via matched-filtering (dotted lines) can be resolved using sparse reconstruction techniques (circles). The dotted horizontal line $h_{{sf}}$ indicates the LASSO trade-off parameter that was used to recover the targets. More details of this simulation can be found in Section~\ref{sec:sims}. (b) Similarly, a target that is masked by the sidelobes of another target (due to the matched-filtering process) can be recovered using LASSO.}
	\vspace{-5mm}
	\label{fig:maskingres}
\end{figure}

Radar detection schemes based on the Neyman-Pearson criterion have been well-studied in the radar community~\cite{Richards2005}. 
However, such schemes are optimal only for scenarios where there is a single target in the radar scene. 
Their adaptation to multi-target scenarios gives rise to issues such as the masking of weak targets by strong ones and limitations of resolution of nearby targets. 
These issues are illustrated in Figure~\ref{fig:maskingres}. 

On the other hand, the radar scene is typically sparse and this prior knowledge promotes the use of sparse reconstruction techniques that have developed rapidly in the last decade. 
Indeed, many recent papers~\cite{Anitori2012,Strohmer,Herman2009,Pribic2012,Strohmera} dealing with sparsity-inspired radar systems have embraced these techniques and have distinguished many interesting advantages of exploiting sparsity in the radar scene. 
In particular, several papers have began examining the use of these sparse reconstruction techniques as a framework for multi-target detection~\cite{Strohmer,Strohmera}. 
Indeed, Figure~\ref{fig:maskingres} shows that using sparse reconstruction techniques allow us to overcome the multi-target issues described above. 

In this paper, we develop a framework to discuss multi-target detection schemes with sparse reconstruction techniques.
We shall build our framework using the Neyman-Pearson criterion where target(s) detection probability is discussed while the probability of false alarm is fixed at a certain level. 
We will also describe an initial result in this framework with LASSO as the sparse reconstruction technique, and we will describe several simulations that validate this result. 
Finally, we discuss several research avenues to further pursue this framework. 

\section{Background}

Let us present some notations that will be used throughout the paper. 
First for any vector $z = [z(1), z(2), \cdots, z(N)] \in \reals^N$, define $S(z)$ as the support of $z$, i.e., $S(z) := \{i \;|\; z(i) \neq 0\}$. 
For any support $S \subset [1, N]$, let $S^c := [1, N]-S$ be the complement of $S$. 
Second, let $\|A\|_\infty$ be the $\ell_\infty$ matrix norm for a matrix $A$ defined as $\|A\|_\infty := \max_p \sum_q |A_{pq}|$ where $A_{pq}$ is the $(p,q)$-th entry of $A$. 
Finally for a $M \times N$ matrix $A$ and support $S$ of size $K$, let $A_S$ be the $M \times K$ matrix whose columns are those corresponding to support $S$. 

\subsection{Traditional Radar Detection}

\label{sec:traditional}

Consider a radar that sends out a waveform $a(t)$ (e.g., a linear chirp) to probe a target scene containing $K$ targets. 
The digital output of the radar (typically sampled at the Nyquist rate of the waveform $a(t)$) can then be modeled by $y = Ax^{*} + e$. 
The vector $y \in \reals^M$ models\footnote{
In this paper, we consider sampling real-valued signals (as opposed to the typical complex-valued signals in radar). 
This choice is for the ease of analysis in the later sections and can be extended to complex-valued signals as well. 
} $M$ consecutive samples of the radar output (also known as \emph{fast-time samples}).\footnote{
If we have a receiving array, then $y \in \reals^M$ represents the sampled output of each array element stacked together into a vector.
}  
The matrix $A = [a_1 | \cdots | a_N] \in \reals^{M \times N}$ has columns $a_i$ that represent the returns of the waveform $a(t)$ (after digital sampling) due to a target at a certain range, Doppler, and angle from the radar (or collectively called the \emph{parameters} of the radar system). 
The vector $x^{*} \in \reals^N$ models the target scene where each of its entries represents the existence of a target with a certain radar parameters (i.e., range, Doppler, angle). 
In our scenario, this vector has $K$ non-zero components and the values of the non-zero components reflect the returned power level of the targets. 
Note that this model assumes that the $K$ targets fall onto a predefined parameter grid of the target scene. 
Finally, the vector $e \in \reals^M$ models the noise at the radar system. 
For this paper, we consider $e$ to be a vector of i.i.d.\ subgaussian random variables with mean 0 and parameter $\sigma$.\footnote{A zero-mean random variable $Z$ is subgaussian if there is a parameter $\sigma$ such that $\E{\exp(tZ)}{} \le \exp(\sigma^2 t^2/2)$ for all $t \in \reals$~\cite{Vershynin2011}. From this definition, a zero-mean Gaussian random variable with variance $\sigma^2$ is a subgaussian random variable with parameter $\sigma$. Examples of subgaussian random variables include Gaussian, Bernoulli, and uniform random variables.} 

Traditional radar detection starts by considering two hypotheses: $\mathcal{H}_1 : y = x(i)^{*} a_i + e$ and $\mathcal{H}_0 : y = e$,
i.e., whether the target scene contains a \emph{single} target generating the returned waveform $a_i$ with returned power $x^{*}(i)$ (hypothesis $\mathcal{H}_1$), or whether the scene has no targets at all (hypothesis $\mathcal{H}_0$). 
The sufficient statistics (derived from the likelihood ratio between the probability density function of $y$ under the two hypotheses) that is used to differentiate between the hypotheses is the matched filter of the radar output $b(i) = a_i^T y$. 
Then, the Neyman-Pearson lemma (or Neyman-Pearson criterion)~\cite{Richards2005} shows that comparing $b(i)$ to a threshold $\eta$ that is set to maintain a fixed probability of false alarm $\alpha$ (i.e., $P_{fa} := \Proba{|b(i)| > \eta | \mathcal{H}_0}{e} = \alpha$) is the \emph{most powerful test} of size $\alpha$. 

While this detection scheme is based on the binary hypotheses of whether or not a \emph{single} target exists in the radar scene, the comparison of the matched-filter output to a threshold is performed even when there are \emph{multiple} targets in the scene. 
To be precise, the output of the radar is first put through a matched filter giving $b = A^T y$ and each matched-filter output $b(i)$ is then compared to the same threshold $\eta$ that is chosen to maintain a fixed false alarm rate $\alpha$. 
Unfortunately, such adaptations result in non-optimal schemes giving rise to issues such as the masking of weak targets by strong ones and limitations of resolution of nearby targets as shown in Figure~\ref{fig:maskingres}.



\subsection{Sparse Reconstruction and Support Recovery}

As the radar scene $x^{*}$ is $K$-sparse (assuming that the targets fall onto the parameter grid), many papers have started promoting the use of sparse reconstruction techniques to recover the radar scene (see e.g.,\cite{Anitori2012,Strohmer,Herman2009,Pribic2012,Strohmera}). 
One of the more well-studied sparse reconstruction technique is the LASSO which, given vector $y$ and matrix $A$, returns a vector $x$ such that
\begin{eqnarray}
	x := \arg \min_{x \in \reals^{N}} \left\{\frac{1}{2} \|y - Ax\|_2^2 + h \|x\|_1 \right\}.
\label{eq:LASSO}
\end{eqnarray}
We note that other sparse reconstruction techniques exist, e.g., CoSAMP~\cite{needell2009cosamp} and complex fast Laplace~\cite{Pribic2012}. 

Conditions for recovering the support $S$ of a $K$-sparse vector $x^{*}$ using LASSO has been analyzed in the literature~\cite{Wainwright2009,Candes2009}. 
Essentially, these papers study the probability of recovering the support $S$ given the trade-off parameter $h$ (see~\eqref{eq:LASSO}), the matrix $A$, the noise level $\sigma$, and signal strength $|x^{*}(i)|$ for $i \in S$. 
These results strengthen our intuition about the trade-off parameter $h$, where a higher $h$ will enforce the greater sparsity of the signal $x$ (determined by the $\ell_1$-norm in~\eqref{eq:LASSO}) while a lower $h$ will enforce greater measurement fidelity (determined by the $\ell_2$-norm in~\eqref{eq:LASSO}).

Similar to this paper, several papers in the literature have analyzed the use of LASSO as a target detection scheme~\cite{Anitori2012,Strohmer,Strohmera}. 
In particular, the results in~\cite{Strohmer,Strohmera} are most closely related to this paper in that they analyze LASSO as a \emph{multiple} target detection scheme (together with a least-squares estimation in a subsequent post-processing step for accurate parameter estimation). 
However, the results therein use slightly different tools (resulting in the limitation of the classes of measurement matrices considered and the introduction of stochastic conditions for the target scene) and does not consider the Neyman-Pearson criterion of fixing false alarms to low level. 

\section{Multi-target Radar Detection}

\subsection{Multi-target Detection Framework}

In this paper, we develop a framework to discuss multi-target radar detection schemes with sparse reconstruction tools. 
These tools are also referred to as Sparse Signal Processing (SSP) in~\cite{Pribic2012} in contrast to traditional radar signal processing. 
The detection schemes considered shall be based on Neyman-Pearson criterion where the probability of false alarm is fixed at a certain level. 

To begin, we elaborate on how false alarms and detection probability can be best discussed in a multi-target detection setting. 
First since the whole target scene is recovered at once with SSP, we propose that the \emph{number of false alarms} $L_{fa}$ (out of $N$) be used to quantify the equivalent type-1 error (false alarm) in our setting. 
Given the support $S$ of the original $K$-sparse scene $x^{*}$, the number of false alarms $L_{fa}$ is the number of non-zero entries outside of $S$ in the recovered vector $x$. 
In other words, $L_{fa} := \left| \left\{ i \;|\; x(i) \neq 0, i \in S^c \right\}\right|$. 
Denoting $P_{fa}$ as the probability of false alarm in a single parameter cell using a traditional detection scheme (i.e., simple thresholding), then $L_{fa}$ is related to how $P_{fa}$ are empirically estimated in a radar system. 
For this, one would typically count the number of threshold crossings outside of true target crossings and then divide this number by the length of the target scene to get an empirical $P_{fa}$. 
For traditional detection schemes, it can be shown that $L_{fa}$ follows a binomial distribution $B(N,P_{fa})$ (assuming that the noise is independent across the parameter cells after matched-filtering).

Second, detection probability can be defined on each of the entries in the support $S$ of $x^{*}$. 
We let $P_D(i)$ for $i \in S$ be the detection probability of the $i$-th target, i.e., $P_D(i) := \Proba{|x(i)| > 0}{}$. 
We note that this is similar to how detection probability is traditionally defined where detection is declared at cell $i$ if $|b(i)| > \eta$ (instead of $|x(i)| > 0$ here).\footnote{We note that this difference is to accommodate for the bias in the estimate of LASSO~\cite{Strohmer,Candes2009}.} 

With false alarms and detection probability thus defined, the aim of our research is two-fold:
\begin{enumerate}
\item First, we are interested in understanding how to tune parameters of the recovery technique (e.g., $h$ in the LASSO) to keep number of false alarms $L_{fa}$ fixed, and
\item based on this false alarm level, we are interested in determining the probability of detecting targets in the scene (i.e., determining $P_D(i)$ for all $i \in S$). 
\end{enumerate}

\subsection{Initial Result}

We shall focus on using LASSO as our sparse reconstruction technique for SSP. 
Our initial result (established in Thm~\ref{thm:main} below) describes how the trade-off parameter $h$ should be set to guarantee (with high probability) that no false alarms appear at the output of LASSO (i.e., $L_{fa} = 0$).

\begin{thm}
\label{thm:main}
Suppose we have radar measurements $y = Ax^{*} + e$ as described in Section~\ref{sec:traditional}. 
Let $S := S(x^{*})$ be the support of $x^{*}$ and define an \emph{incoherence} parameter $\gamma = \gamma(x^{*}, A) \in (0, 1]$ as
\begin{eqnarray}
	\gamma := \left\| A_{S^c}^T A_S \left(A_S^T A_S\right)^{-1}\right\|_\infty. 
	\label{eq:gamma}
\end{eqnarray}

Fix a failure probability $p$. If $A_S^T A_S$ is invertible and if $x$ is the LASSO solution with parameter $h$ given by
\begin{eqnarray}
	h \ge \frac{\sqrt{2} \sigma \max_{j \in S^c} \|a_j\|_2}{1 - \gamma} \sqrt{\ln\left( \frac{2(N-K)}{p} \right)},
	\label{eq:h_param}
\end{eqnarray}
then 
	$\Proba{L_{fa} > 0}{e} = \Proba{S(x) \not \subset S(x^{*})}{e} \le p$. 

\end{thm}

The failure probability $p$ 
is the \emph{maximum} probability of getting \emph{at least} one false alarm in the recovered vector $x$ for a particular draw of the noise vector $e$. 
Thus to minimize the appearance of \emph{any} false alarms, one would set $p$ to be as small as possible. 

The setting of the failure probability $p$ affects the value of $h$ in LASSO. 
One can see that setting a smaller $p$ would make the required $h$ larger (though only logarithmically). 
This is logical since the way to reduce false alarms in the recovered vector would be to enforce more sparsity in the recovery. 

One could also note that the form that the required $h$ takes in~\eqref{eq:h_param} is similar to the threshold value that one sets for a traditional radar detector. 
A traditional detector threshold (for parameter cell $i$ under subgaussian noise assumptions) is typically set at $\eta \ge \|a_i\|_2 \sigma \sqrt{\ln(1/P_{fa})}$. 
By comparing $h$ to $\eta$, we observe that both values need to be set above the noise level $\sigma$. 
We can also remark that the failure probability $p$ (for the whole parameter space) in $h$ can be compared to the probability of false alarm $P_{fa}$ (for a single parameter cell under test) in $\eta$. 
Finally, we remark that the additional terms appearing in $h$ reflect the fact that the whole target scene is recovered at once for SSP. 
These terms include the logarithmic dependence on $N-K$, the maximum returned signal strength $\max_{j \in S^c} \|a_j\|_2$, and the incoherence term $\gamma$. 

Let us elaborate more on the incoherence term $\gamma$. 
Suppose $A_S^T A_S$ is identity and suppose the basis vectors $a_j$ outside of the support $S$ are incoherent with the ones in the support (i.e., $a_j^T a_i$ is small for $i \in S$ and $j \in S^{c}$). 
Then $\gamma$ will be small and the dependence on this incoherency will go away in~\eqref{eq:h_param}. 
This is reasonable as the more incoherent the basis vectors outside the support are, the less they will be confused for target basis vectors under the influence of noise. 
However when $\gamma$ is large, $h$ can also become large. 
As we shall see in Section~\ref{sec:sims}, a large $h$ has a negative effect on the probability of detection of targets. 

This incoherence factor $\gamma$ can be bounded by more familiar qualities of measurements matrices $A$ such as the RIP. 
For example by combining Prop. 7.2 and Prop. 2.5 in~\cite{Rauhut2010b} and supposing that the matrix $A$ satisfy $K$-RIP with conditioning $\delta$, we see that $\gamma \le \frac{\sqrt{K}\delta}{1 - \delta}$. 
Thus if the RIP conditioning $\delta$ is kept small, then the incoherence parameter $\gamma$ will also be small. 
We note that the RIP of the matrix $A$ also implies that $A_S^T A_S$ is invertible. 

The proof of this result follows closely the proof of~\cite[Thm 1]{Wainwright2009} but is left out due to space constraints. 
We reiterate the fact that this result is simply an initial step towards our goal of establishing a multi-target detector. 
Its various shortcomings and the future steps towards our multi-target detector goal will be described in Section~\ref{sec:conclusion}.

\section{Simulations}
\label{sec:sims}

We perform some simulations to validate our theory of how the parameter $h$ of LASSO controls the (maximum) probability $p$ of having at least one false alarm. 
To keep the simulations simple, we assume a Nyquist-rate, uniformly-sampled, range-only pulse radar setup. 
In this setup, the length-$M$ measurements $y$ are taken over a single pulse repetition time, the length-$N$ vector $x^{*}$ represents an unknown range profile with $K$ targets, the receiver noise $e$ is assumed i.i.d.\ (complex) Gaussian with zero mean and variance $\sigma^2$, and the $M \times N$ measurement matrix $A$ contains delayed replicas of a transmitted (complex) linear chirp waveform (with a length of 25 time units and with bandwidth equal to the unit sampling frequency). 
In the simulations, the columns of $A$ are normalized to norm 1. 
This means that pulse compression gain would have been taken into account when considering the SNR values of the targets appearing in $x^{*}$. 

To obtain a super-resolution effect, the range estimation grid is upsampled to $N = 250$ points from $M = 108$ measurements. 
The number of targets $K$ is varied from 1 to 3 and their locations are set at positions 100, 104 and 133 on the upsampled grid. 
As an example, the crosses in Figure~\ref{fig:maskingres}(a) shows the positions of $K = 2$ targets in the radar scene appearing at positions 100 and 104. 

The incoherence parameters calculated for $K = 1,2,3$ described above correspond to $\gamma = 0.8272, 0.8332, 0.8338$ respectively. 
These high $\gamma$ values for the linear chirp waveform implies that the recovery of the target scene with $L_{fa} = 0$ requires a high value of $h$. 
We shall see later than a large $h$ reduces the probability of detection of low SNR targets. 

In the simulations that follow, the amplitude of each target in $x^{*}$ is set to 1 while the noise variance $\sigma^2$ is set to achieve a certain target SNR, i.e. $\sigma^2 = \frac{1}{\mbox{SNR}}$. 
For each value of $K$, the following values of SNR are used (in dB): $19, 22, 25, 28, 31, 33$. 
By setting the failure probability $p = 0.1$, Table~\ref{table:hvalues} shows the minimum values of the $h$ parameter given in~\eqref{eq:h_param}. 

\begin{table}
{\footnotesize
\begin{tabular}{c|cccccc}
SNR (dB) & 19 & 22 & 25 & 28 & 31 & 33 \\ \hline
$K = 1$ & 2.6798 & 1.8972 & 1.3431 & 0.9508 & 0.6731 & 0.5347 \\
$K = 2$ & 2.7754 & 1.9648 & 1.3910 & 0.9847 & 0.6971 & 0.5538 \\
$K = 3$ & 2.7843 & 1.9712 & 1.3955 & 0.9879 & 0.6994 & 0.5555 
\end{tabular}
}
\caption{\small Minimum values of $h$ calculated from~\eqref{eq:h_param}.}
\label{table:hvalues} 
\end{table}

For each value of $K$ and SNR, 100 instances of the noise vector $e$ is drawn. 
Then for each noise instance, the problem $y = Ax^{*} + e$ is formulated and LASSO is run with the $h$ parameter set in Table~\ref{table:hvalues}.\footnote{CVX is used to solve LASSO~\cite{gb08,cvx}. Additionally, recovered values below $h/10$ for the values of $h$ in Table~\ref{table:hvalues} are set to $0$.} 
One such run of the LASSO for $K = 2$ and $\mbox{SNR} = 33 \mbox{dB}$ is shown in Figure~\ref{fig:maskingres}(a). 
In that example, the two targets (crosses) that are close together in range which are initially unresolved via matched-filtering (dotted lines) can be resolved using LASSO (circles). 

\begin{figure}[t]
	\begin{center}
\includegraphics[width=8cm]{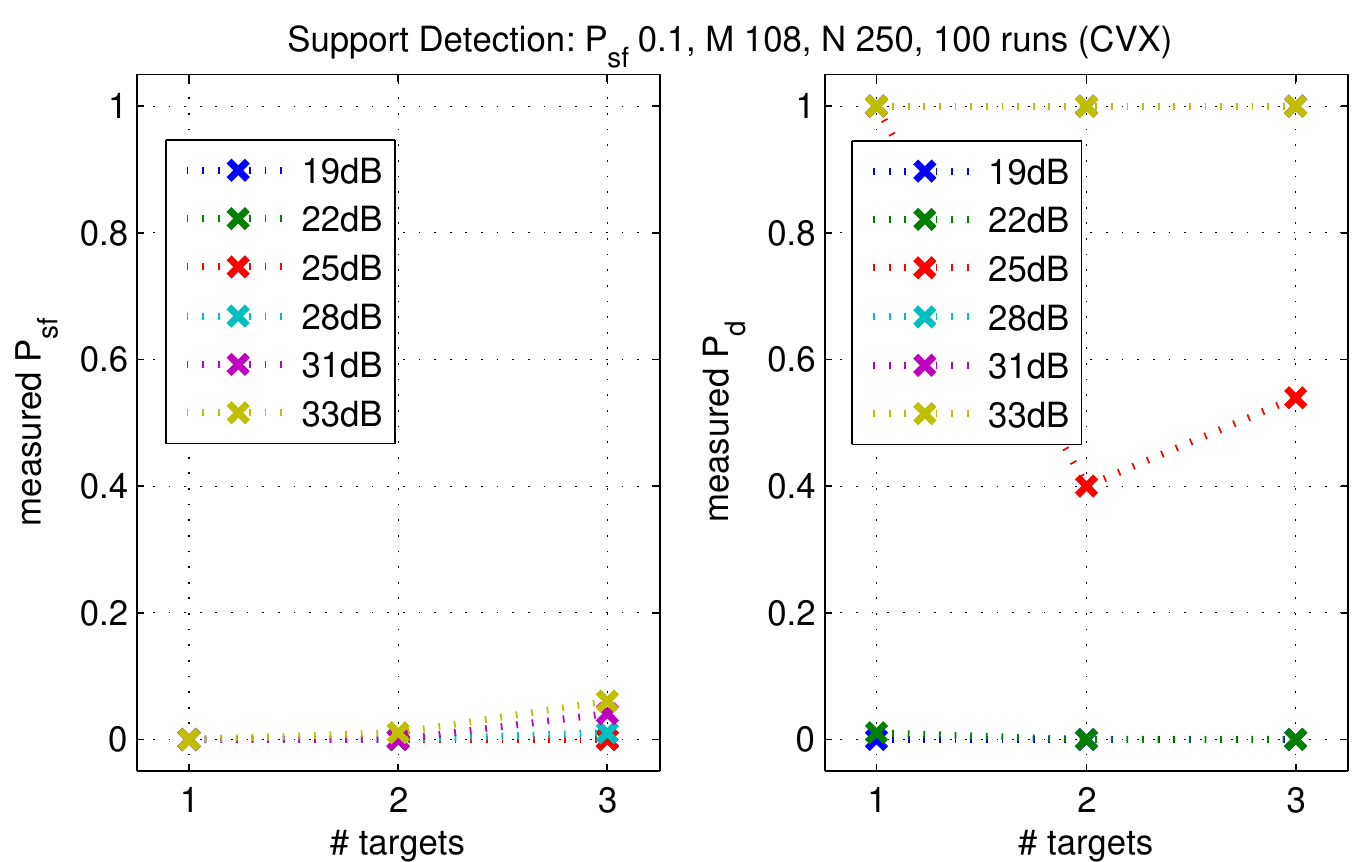} 
	\end{center}
	\vspace{-5mm}
	\caption{\small Empirical failure probability $p$ (or $P_{sf}$ as indicated in the figure) and empirical probability of detection of all targets (i.e., $K^{-1} \sum_{i=1}^K P_D(i)$) calculated for each $K$ and SNR over the 100 instances of the noise vector $e$.}
	\vspace{-5mm}
	\label{fig:sims}
\end{figure}

Figure~\ref{fig:sims} shows the empirical failure probability $p$ and the empirical probability of detection of \emph{all} targets (i.e., $K^{-1} \sum_{i=1}^K P_D(i)$) calculated over all the runs. 
First, we observe that the failure probability is indeed below the preset value of $p = 0.1$ as expected. 
Second, we observe that the probability of detecting targets depends on whether the target power (set to 1 here) exceeds the parameter $h$ value. 
For values of $h$ that are much greater 1 (corresponding to $\mbox{SNR} = 19, 22$ from Table~\ref{table:hvalues}), the empirical probability of detection is $0$. 
Similarly, for values of $h$ below 1 (corresponding to $\mbox{SNR} = 28, 31, 33$ from Table~\ref{table:hvalues}), the empirical probability of detection raises to $1$. 
When $h$ takes value close to but greater than 1 (corresponding to $\mbox{SNR} = 25$), the empirical probability of detection ranges between $0$ and $1$. 

\section{Future Work}
\label{sec:conclusion}

First, our result can be extended to complex signals to match typical radar problems. 
A similar extension has been done in~\cite{Strohmer}. 

Second, the correspondence of $h$ to the probability of detection $P_D(i)$ of a target (using LASSO) as observed in Section~\ref{sec:sims} can be made using tools from~\cite{Wainwright2009}. 
On-going research efforts towards this goal is being made. 
Similar work on this topic (using different tools) has been described in~\cite{Strohmer,Strohmera}. 

Third, we have consider only target detection under subgaussian noise in the current paper. 
Generalization of the noise model to include various types of clutter statistics will make the result useful for the detection of targets in clutter. 

Fourth, our initial result only describes condition for $L_{fa} = 0$. 
In analogy to traditional detection schemes, one can imagine that by allowing the appearance of some false alarms (i.e., for $L_{fa} > 0$), the detection probability of targets can be improved. 

Lastly, we have only considered point targets that lie on the parameter grid. 
As most real targets are extended and do not lie on the grid, considerations for the recovery of such targets need to be made. 
Indeed, various novel sparsity-based techniques that deal with off-the-grid and extended targets have already appeared in the literature (see e.g.,~\cite{Radmila12}). 

\bibliographystyle{IEEEtran}
\bibliography{bibliography}



\end{document}